\documentclass{aastex}          
\usepackage{spr-astr-addons}    
\usepackage{natbib}
\usepackage{graphicx}
\usepackage{multirow}
\usepackage{txfonts}

\begin{document}
\shorttitle{Disk-Jet Connections of MAXI~J1836-194 with TCAF Solution}
\shortauthors{Jana, Debnath, Chakrabarti \& Chatterjee}
\title{Inference on Disk-Jet Connection of MAXI~J1836-194: Analysis with the TCAF Solution}
\author{Arghajit Jana\altaffilmark{1}, Dipak Debnath\altaffilmark{1}, Sandip K. Chakrabarti\altaffilmark{1,2},
Debjit Chatterjee\altaffilmark{1}}
\altaffiltext{1}{Indian Centre for Space Physics, 43 Chalantika, Garia St. Rd., Kolkata, 700084, India.}
\altaffiltext{2}{S. N. Bose National Centre for Basic Sciences, Salt Lake, Kolkata, 700106, India.}
\email{argha0004@gmail.com; dipakcsp@gmail.com; chakraba@bose.res.in; debjitchatterjee92@gmail.com}


\begin{abstract}
Galactic transient black hole candidate (BHC) MAXI~J1836-194 was discovered on 2011 Aug 30, by MAXI/GSC and Swift/BAT. 
The source activity during this outburst was continued for $\sim 3$ months before entering into the quiescent state. 
It again became active in March 2012 and continued for another $\sim 2$ months. In this paper, $3-25$ $keV$ RXTE/PCA 
spectra of 2011 outburst and $0.5-10.0$ $keV$ Swift/XRT data during its 2012 outburst are analyzed with the two-component
advective flow (TCAF) model based fits file in XSPEC. We calculate the X-ray contributions coming from jets/outflow 
using a newly developed method based on the deviation of the TCAF model normalization. We also studied the correlation 
between observed radio and estimated jet X-ray fluxes. The correlation indices ($b$) are found to be $1.79$, and $0.61$, 
when the $7.45$~GHz VLA radio flux is correlated with the total X-ray and the jet X-ray fluxes in $3-25$~keV range 
respectively. It has been found that the jet contributes in X-rays up to a maximum of 86\% during its 2011 outburst.
This makes the BHC MAXI~J1836-194 to be strongly jet dominated during the initial rising phase.

\end{abstract}


\section{Introduction}  
Jets and outflows are very common in the active galactic nuclei (AGN). It is also observed in many 
Galactic black hole candidates (BHCs), such as, GRS~1758-258 (Rodr\'iguez et al. 1992), 1E~1740.7-2942
(Mirabel et al. 1992), Cyg X-1 (Stirling et al. 2001) etc. In general, two types of jets are observed
in Galactic BHCs: continuous or compact jet and discrete or blobby jet (see, Chakrabarti \& Nandi, 2000
for more details). The exact mechanism for the production of jets is still unclear. In the literature,
several models have been put forward to explain jet and outflows (Blandford \& Znajek 1977; Blandford
\& Payne 1982; Chakrabarti \& Bhaskaran 1992). In general, the magnetic field is considered to be the 
reason behind the collimation of jets (Camenzind 1989)

Chakrabarti (1999a,b; hereafter C99a and C99b respectively), Das \& Chakrabarti (1999) have calculated
the outflow rate from the inflow accretion rate using hydrodynamics of in-falling and outgoing transonic 
flows. They show that the thermal pressure could be sufficient to supply matter to the outflows and 
accelerate them from low to moderate Lorentz factors. Transonic flow solutions naturally connect the disk
and jet as they belong to the same class of solutions with opposite boundary conditions in the sense that
an accretion flow is subsonic at infinity and supersonic at the black hole horizon while the jet is subsonic
close to the horizon and supersonic at infinity. Using the properties of the transonic flows, the two-component
advective flow (TCAF) solution (Chakrabarti \& Titarchuk 1995; Chakrabarti 1997, and references therein)
for the accretion flow around a black hole was proposed. In this solution, an accretion disk has two components
of the inflowing matter: a low viscous, low angular momentum sub-Keplerian halo component surrounds a high
viscous, high angular momentum Keplerian disk component. The sub-Keplerian flow forms an axisymmetric shock
at the centrifugal barrier and the inflowing matter slows down at the shock location ($X_s$) making the region
hot and puffed-up. This post-shock region is known as the CENBOL or CENtrifugal pressure supported BOundary
Layer. It acts as the so-called `corona' or `Compton cloud' (Sunyaev \& Titarchuk 1980, 1985). Low energy 
photons coming from the Keplerian disk are inverse-Comptonized at CENBOL and become hard photons. They create
the power-law (PL) component of the observed spectrum of black holes. Observed multi-color blackbody i.e., disk
blackbody (DBB) component is due to thermal photons originated from the Keplerian disk (similar to a truncated
Shakura \& Sunyaev disk, but the temperature distribution is modified by reflected hard photons).

In the TCAF solution, the CENBOL is also the base of the jet (C99a,b). Mass outflow rate depends on the accretion
rate, shock radius and the shock compression ratio ($R$), which is the ratio between post- and pre-shock densities.
Matters are driven outward due to thermal pressure gradient force. The outflowing matter or jet moves slowly up 
to the sonic surface ($\sim 2.5 X_s$, C99a,b). After that, they move away supersonically. Due to the adiabatic
expansion, temperature falls as the jet moves away. It emits electromagnetic radiation in all wavebands. The 
theoretical dependence of the ratio of outflow and inflow rates on the compression ratio suggests that the outflow
rate is not maximum in the hard state. Rather, it is maximum in the intermediate states when the shock strength is 
intermediate (C99a,b). In the intermediate states, matter supply from the companion is much higher than that in 
the hard state. As the Keplerian rate is increased, the CENBOL is cooled down and its size is reduced. When the 
jet-base is cooled, the flow suddenly becomes supersonic, separating it from the CENBOL to produce blobby jets
(Chakrabarti, 1999b; Das \& Chakrabarti 1999). The reduction of the outflow rate on the Keplerian disk rate has 
been verified by numerical simulations (Garain et al. 2012). In the soft state, the Keplerian rate becomes very 
high and completely cools down the CENBOL, quenching the jet altogether.

Recently, TCAF solution has been successfully implemented as an additive table model into HEASARC's spectral
analysis software XSPEC (Arnaud 1996) to fit BH spectrum (Debnath et al. 2014, 2015a). Using the TCAF model 
fitted spectral analysis, accretion dynamics of several black holes have been explained satisfactorily (Mondal 
et al. 2014, 2016; Debnath et al. 2015b, 2017; Chatterjee et al. 2016, 2018; Jana et al. 2016, hereafter 
Paper-I; Bhattacharjee et al. 2017; Molla et al. 2017). To fit a BH spectrum with the TCAF model based {\it fits} 
file in XSPEC, one needs to supply two types of accretion rates (Keplerian $\dot{m_d}$, and sub-Keplerian 
$\dot{m_h}$) in units of Eddington rate, shock location ($X_s$) in Schwarzschild radius ($r_s=2 G M_{BH}/c^2$),
and shock compression ratio ($R$), if the mass ($M_{BH}$ in $M_\odot$) and normalization are known. If the 
mass is unknown, one can also estimate it from the spectral analysis with the current version of the TCAF 
model {\it fits} file (Molla et al. 2016, 2017; Chatterjee et al. 2016; Paper-I; Debnath et al. 2017). 

Unlike other models, TCAF model normalization $N$ is constant across the spectral states for a particular
BHC observed by a given instrument. Using this, Jana et al. (2017, hereafter JCD17) separated the X-ray 
contribution from the accretion disk or inflowing matter ($F_{inf}$) from jets or outflowing matter ($F_{ouf}$). 
In JCD17, estimation of the jet X-ray flux ($F_{ouf}$) and its properties during the 2005 outburst of 
BHC Swift~J1753.5-0127 are studied. In the current paper, using the same method, we separate X-ray contribution
from jets/outflow for the Galactic BHC MAXI~J1836-194 during its 2011 and 2012 outbursts. The properties of
X-ray jets are also studied during the outbursts as well as the intervening quiescent phase.

Galactic transient BHC MAXI~J1836-194 was discovered on 29 Aug 2011 by MAXI/GSC (Negoro et al. 2011). Swift/BAT also 
observed it simultaneously at R.A. $= 18^h35^m43^s.43$, Dec $= -19^\circ 19'12''.1$.
During this outburst epoch, the source was active for $\sim 3$ months before going to the quiescent state. It again
showed a short activity on March 2012 (Krimm et al. 2012; Yang et al. 2012a). This source was studied extensively in
multi wave-bands: from radio, optical to X-rays (Ferrigno et al. 2012; Reis et al. 2012; Yang et al. 2012b; Russell 
et al. 2013, 2014ab, 2015; Paper-I) during its 2011 outburst. It has a short orbital period of $< 4.9$~hrs and a low
inclination angle ($4-15^\circ$; Russell et al. 2014a). This BHC is rapidly spinning with a spin parameter of 
$a=0.88\pm0.03$ (Ries et al. 2012). Russell et al. (2014a) reported the mass of the black hole to be $>1.9~M_\odot$, 
if the source is located at $4$~kpc and $>7~M_\odot$ if the distance is $10$~kpc. In Paper-I, using TCAF model fitted
spectra, Jana et al. suggested the mass of the BH to be in between $7.5-11~M_\odot$ or, more precisely, 
$9.5^{+1.5}_{-2}$. Russell et al. (2014a) also suggested that the binary companion could be a low mass
($<0.65~M_\odot$) star.

Our paper is organized in the following way: In \S2, we briefly discuss about observation and data analysis methods 
for estimation of the jet X-ray fluxes. In \S3, we discuss results based on our analysis which includes the study 
of accretion flow properties of the source during its 2011 \& 2012 outbursts and the intervening quiescent phase. 
The evolution of the X-ray jet and its correlation with the observed radio flux density are discussed. Finally,
in \S5, we make concluding remarks.

\section{Observation, Data Analysis and Method of Estimating Jet X-ray Flux}

Jana et al. (2016) studied spectral and timing properties of MAXI~J1836-194 during its 2011 outburst in details. 
Here, to estimate jet X-ray fluxes from the spectral analysis with the TCAF model, we use $3-25$ $keV$ RXTE
proportional counter unit 2 (PCU2) data of total $35$ observational IDs (starting from the first PCA observation
day, i.e., 31 August 2011 or Modified Julian Day, i.e., MJD=55804 till 24 November 2011 or MJD = 55889). For the 
data extraction and analysis, we follow the same method as in Paper-I. After spending $\sim 4$ months in the 
quiescent state, MAXI~J1836-194 showed renewed activity in March 2012. We also analyzed $\sim 15$ Swift/XRT 
observations between 2012 March 12 and May 1. We use standard `$xrtpipeline$' command to extract $lighcurve$,
$.pha$ and $.arf$ files. $0.5-10$ $keV$ XRT spectra are fitted with the TCAF model. However, due to lack of data
points with low signal-to-noise ratio, acceptable $\chi^2$ statistics is obtained only for two observations during
the 2012 outburst.

While fitting, a model normalization is used as a multiplicative `factor' that converts the observed spectra
to match the theoretical model spectra. In general, one may require different normalization for different 
observations. In TCAF, however, the entire spectrum is an outcome of the solution and thus the model normalization
`N' only depends on the intrinsic source parameters, namely the mass of the BH, the distance of the source and the 
disk inclination angle. Thus it must remain a constant for a source observed by a particular satellite instrument. 
In our fit, one may still see some deviation of N, if the data is not of uniform quality, or, if the disk precesses 
(i.e., if the effective disk area changes) or dominance of other physical processes such as X-rays from jets or 
outflows, which are not included in the current version of the theoretical model {\it fits} file. For instance,
if a jet is present and the base contributes to the observed X-rays, we require a higher value of $N$ while fitting
spectrum with the TCAF model {\it fits} file. This, together with a simultaneous observation of activities in radio
can confirm if the base of the jet is active in X-rays. While fitting the $3-25$~keV RXTE/PCA data of the 2011 
outburst of MAXI~J1836-194, we did not obtain a constant $N$ value for all observations. The normalization $N$ 
generally varied within a narrow range of $0.25-0.35$ when radio is not very strong. However, in some observations,
we require higher values of $N$, when the jet is also found to be stronger, i.e., observed radio flux density is 
high. When the jet is active, its contribution to X-rays also increases. Thus, the total X-ray flux ($F_X$) obtained 
from $3-25$ $keV$ RXTE/PCA data is the sum total of the contribution both from the jets as well as from the accretion
disk. On 2011 Oct. 22 (MJD=55856), $N$ was found to be $0.25$ which is the lowest value. This leads us to assume that
on this date, the X-ray flux is completely from the accretion disk (see, JDC17). To estimate the X-ray contribution
only from the accretion disk or inflowing matter ($F_{inf}$), we refit all the spectra with $N$ frozen at $0.25$. By 
taking the differences between $F_X$  (which is the flux in $3-25$~keV of our previous model fitted spectra, when all
model parameters are kept free) and $F_{inf}$ (which is the flux of our later fits with fixed $N=0.25$) we can
estimate the jet X-ray flux ($F_{ouf}$) to be given by,  
$$
F_{ouf}=F_X - F_{inf}. \eqno{(1)}
$$

We have also calculated $F_{ouf}$ during the 2012 outburst based on Swift/XRT analysis. From RXTE/PCA spectral
analysis of the 2011 outburst, $N$ was found to be minimum, $\sim 0.25-0.253$ during MJD= 55850 to 55867 during 
the declining phase of the hard state. This means that the X-ray jet must be lowest or inactive in these days 
(JCD17). $0.5-10.0$~keV Swift/XRT spectrum of 2011 Oct. 25 (MJD=55859) is now fitted with the TCAF model {\it fits}
file to have an estimation of the model normalization at low or no X-ray jet condition for XRT spectrum in the 
specified energy band. For the best model fit, we find $N=34.28$ (see Table 1). We also checked other few XRT
observations around 2011 Oct. 22 (MJD=55856, with the minimum value of $N=0.25$ in $3-25$~keV PCA data), and 
found similar $N$ values for the XRT. Thus we used this Swift/XRT spectrum fitted $N$ value (as no X-ray jet 
`$N$' for XRT in $0.5-10.0$~keV for MAXI~J1836-194) to calculate the jet X-ray flux for the 2012 outburst of the
source. We freeze mass at $9.54$ $M_{\odot}$ while fitting the XRT data.

\noindent{Note that $F_X$, and $F_{inf}$ fluxes for $3-25$~keV RXTE/PCA spectra are obtained using `flux $3.0$ $25.0$' 
command, after obtaining the best model fits in XSPEC.}

\section{Results}

We study the source during its initial $\sim 10$ months (from 2011 August 31 to 2012 May 13; i.e., MJD=55804
to 56060) period after the discovery on 2011 August 30 MAXI~J1836-194 showed two outbursts in 2011 and 2012 
with duration of $\sim 3$ \& $2$ months respectively separated by $\sim 4$ months of quiescent period, during 
the period of our analysis. We compare variations of X-ray and radio intensities of the source, along with its 
spectral and jet properties. 

\subsection{X-ray and Radio Lightcurves}

We plot $15-50$ $keV$ Swift/BAT and $2-10$ $keV$ MAXI/GSC lightcurves in Fig. 1(a-b). Hardness ratio (HR) of
Swift/BAT and MAXI/GSC count rates are shown in Fig. 1(c). HR is defined as the ratio between $15-50$ $keV$ BAT
and $2-10$ $keV$ GSC count rates. In Fig. 1(d), radio flux densities in $5$ and $7.45$~GHz of VLA, and $5.5$~GHz 
of ATCA data are shown. Radio data are taken from Russell et al. (2014b, 2015) papers.

The X-ray lightcurves are plotted for $\sim 10$ months between 2011 Aug 26 and 2012 Jun 7. In the rising phase
of the 2011 outburst, both MAXI/GSC and Swift/BAT fluxes increased rapidly starting from 2011 Aug 29 (MJD = 55802), 
just one day before its discovery. They attained a maximum peak flux of around Sept 6, 2011 (MJD=55810). After that, 
the flux decreased slowly, although Swift/BAT showed another small peak around Sept 9, 2011 (MJD=55813). Both 2011
\& 2012 outbursts could be termed as `Fast-Rise-Slow-Decay' (FRSD) type as variation shown in the outburst profiles 
(Debnath et al. 2010). From the spectral evolution, it could be termed as `type-II' or `harder type' of BH binaries, 
since it does not show softer spectral states during its outbursts (Debnath et al. 2017). The BHC MAXI~J1836-194 was
active for $\sim 3$ months till MJD $\sim 55890$ during its 2011 outburst. Then it entered in the quiescent state, 
which continued for $\sim 4$ months. The X-ray flux again started to rise around 2012 March 12 (MJD=55998; Krimm 
et al. 2012, Yang et al. 2012). $15-50$~keV Swift/BAT flux increased significantly during this epoch of the 2012 
outburst, although $2-10$~keV MAXI/GSC flux did not show any significant change. The 2012 outburst is much weaker
as compare to the 2011 outburst and it continued for $\sim 2$ months. On 2012 March 24 (MJD=56010), maximum flux 
in Swift/BAT was observed. During this outburst, BAT flux rapidly increased for initial $\sim 20$ days before it 
decreased. After that, it moved to the quiescent phase, where the flux remained almost constant at very low values. 

HR roughly indicates whether a BHC is in the soft state or in the hard state. During the 2011 outburst, HR was 
minimum on MJD $\sim 55820$. The BHC was in the hard-intermediate state during that time (Paper-I). Paper-I also 
showed that the ARRID (accretion rate ratio intensity diagram) is a better alternative to the `q'-diagram or HID
(hardness intensity diagram) as in Ferrigno et al. (2012) for describing transitions between spectral states. 
For a quick look, HR is more useful since to obtain ARRID, one needs to fit spectra with the TCAF model. From 
Fig. 1c, we observe that at the very beginning of the 2011 outburst, HR increased rapidly, before it started to 
decrease slowly until MJD=55820. After that, HR increased slowly and became roughly constant at $\sim 2$. This
trend of higher (hard state) HR value continued more or less in the quiescent state as well as in the 2012 outburst. 
In the quiescent state, HR fluctuated between $2$ and $6$. It indicates that during the quiescent state, the source
was in the hard state with a very low mass accretion rate. During the 2012 outburst, the same HR was observed. 
However, to check if during the entire 2012 outburst period, the object remained in a hard state, we need to
carry out the spectral analysis.

\subsection{Accretion Flow Properties of MAXI~J1836-194} 

We study accretion flow properties of MAXI~J1836-194 during its 2011 \& 2012 outbursts and the intervening
quiescent phase. The detailed spectral and timing analysis using RXTE/PCA data to infer accretion flow dynamics 
of the source during its 2011 outburst has already been reported in Paper-I. Here, we extend our analysis period 
to cover quiescence as well as the 2012 outburst, wherever the data is available. We mainly concentrate on the
properties of the jet in X-rays.

\subsubsection{\textbf{2011 Outburst}}
As stated earlier, the detailed study of the spectral and timing analysis to infer the accretion flow dynamics of
the source during this outburst has already been reported in Paper-I using $2.5-25$ $keV$ RXTE/PCA data. Based on
the variation of accretion rate ratio (ARR=$\dot{m_h}$/$\dot{m_d}$), nature of QPOs, they classified the entire 
outburst into two spectral states: hard (HS) and hard-intermediate (HIMS). These observed states form a hysteresis
loop as: HS (Ris.) $\rightarrow$ HIMS (Ris.) $\rightarrow$ HIMS (Dec.) $\rightarrow$ HS (Dec.). No signature of 
the softer spectral states, i.e., SIMS and SS were observed. Since the source is one of the shorter orbital period
BHCs, there could be high amount of low angular momentum halo component from the companion winds, hardening the flow. 
From the spectral analysis, Jana et al. (2016) estimated the probable mass of BH to be in the range of $7.5-11.0$
$M_\odot$. Taking the average, the probable mass of the BH is about $9.54^{+1.47}_{-2.03}$ $M_\odot$. They also 
estimated the viscous time scale as $\sim10$ days, obtained from the differences in occurrences of the peaks of two 
types of accretion rates ($\dot{m_d}$ \& $\dot{m_h}$). 

While fitting the spectra with the TCAF model {\it fits} file, Jana et al. (2016) did not find the model normalization
($N$) to be roughly constant throughout the outburst. Generally it varied within $0.25-0.35$. However, in some 
observations, very high $N$ values were required to fit the spectra, particularly when the radio flux densities
were high. This indicates that these higher values of $N$ may be due to the excess contribution of the X-rays from the 
jets (JCD17). We use the same method as in JCD17 to obtain the X-ray flux of the jet in the present object.

\subsubsection{\textbf{Quiescent State}}

In general, a BHC is considered to be in quiescent state when the X-ray luminosity $L_X \textless 10^{34}$ $ergs/s$. 
It is believed that the quiescent state is the extended phase of hard/low-hard state with very low accretion rate and 
low luminosity. MAXI~J1836-194 was in the quiescent state for $\sim 4$ months between its 2011 and 2012 outbursts. 
X-ray luminosity in $2-10$~keV band (calculated from MAXI/GSC observed flux) were observed as low as $L_X \sim 10^{33}$
$ergs/s$ during the phase. HR was observed to have higher (in between $\sim 2-6$). The radio jet was observed during
this phase though the luminosity was much lower as compared to the 2011 outburst. Nature of the X-ray flux, observation
of radio flux and higher HR indicate that MAXI~J1836-194 was in hard/low-hard state during this phase with very low 
accretion rate.

\subsubsection{\textbf{2012 Outburst}}

The 2012 outburst is much weaker as compared to the 2011 outburst. Luminosity was about one hundred times lower
than that of the 2011 outburst. This new flaring activity of MAXI~J1836-194 was detected by Swift/BAT on 2012 
March 10 (Krimm et al. 2012). The source was active for about $\sim 60$ days. Grebenev et al. (2013) reported
that the $0.3-400$~keV Swift+INTEGRAL spectra are power-law dominated with very little contributions from the 
disk blackbody component. We have analyzed Swift/XRT spectra in the energy range of $0.5-10.0$~keV with the TCAF
model. Though due to the lack of data points and low signal-to-noise ratio, we are unable to achieve acceptable
$\chi^2$-statistics in many observations with TCAF model or with the phenomenological disk blackbody plus powerlaw
models. We only found better $\chi^2$ statistics in two observations on 2012 March 20 (MJD=56006) and 2012 March
27 (MJD=56013). Model fitted parameters of these two XRT observations are given in Table 1.

The TCAF model fitted extracted values, and high values of ARR suggest that the disk is highly dominated by the
sub-Keplerian halo component. The sub-Keplerian halo rate ($\dot{m_h}$) is found to be much higher as compared 
to the Keplerian disk rate ($\dot{m_d}$). So ARRs are also found to be high ($\sim13-14$). The 2012 outburst was 
dominated by the low viscous sub-Keplerian matter similar to the 2011 outburst of MAXI~J1836-194. So, we can 
assume that the viscosity was lower than the critical value during the entire period of the outburst (Chakrabarti,
1997 and references therein). Due to this, a low supply in Keplerian disk component is observed, which was unable
to cool the CENBOL sufficiently. In our analysis of the two XRT observations, fits are obtained with strong shocks
of compression ratio $R\sim3.5$. Higher values ($74.92$ \& $93.82$) of the TCAF model normalization indicate that
may be the X-ray contribution of the jet was strong as well. Radio observation also supports this. During the 
fitting, we keep mass of the BH frozen at $9.54$~$M_{\odot}$.

\subsection{Disk-Jet Connections}

\subsubsection{\textbf{Evolution of Jets}} 

We calculated X-ray contributions from the jet during the 2011 and the 2012 outbursts using procedure of JCD17.
The variation of total X-ray flux ($F_X$), accretion disk X-ray flux ($F_{inf}$), and jet X-ray flux ($F_{ouf}$) 
are shown in Fig. 2 (a-c). The variation of the TCAF fitted normalization ($N$) is shown in Fig. 2d. $7.45$ $GHz$ 
VLA radio data are plotted in Fig. 2e. The jet X-ray flux ($F_{ouf}$) is found to increase slowly as the outburst
progressed during the 2011 outburst. It attained the maximum value on $\sim$ 2011 Sept. 06 (MJD=55810). VLA first
observed the source in $7.45$~GHz radio band on 2011 Sept. 03 (MJD=55807), with a flux density of $27$~mJy. It was 
roughly constant for the next $\sim 20$~days. The source was in the HIMS during this phase of the outburst. $F_{ouf}$
attained its peak value on 2011 Sept. 09 (MJD=55813). Then it showed roughly constant nature. $F_{ouf}$ started to
decrease after 2011 Sept. 23 (MJD=55827). Radio flux density also showed a similar behavior. The BHC entered into
the HS (Dec.) on 2011 Oct. 01 (MJD=55835), which continued till the end of the outburst. During this phase of the 
outburst, both $F_{ouf}$, and $F_R$ were found to be very low.

TCAF normalization ($N$) also showed a similar behavior as $F_R$ and $F_{ouf}$. It increased slowly in the rising
phase of the 2011 outburst. On MJD=55813.56 and 55818.84, much higher $N$ values of $1.99$, and $2.07$ respectively 
were required to fit the spectra with the TCAF {\it fits} file. After that, $N$ decreased slowly till MJD $\sim$
55830 after which $N$ varied within a narrow range of $\sim 0.25-0.35$. 

Four spectra from different spectral states are shown in Fig. 3(a-d). The spectra are fitted with free (black solid curve)
or frozen (at $0.25$; red dashed curve) normalization values. The jet spectra are obtained by taking differences between 
them and shown with blue dotted-dashed curve. We can see that the jet spectra are stronger in the HIMS and also they are 
harder than the disk spectra.

We have also calculated $0.5-10$~keV X-ray jet flux during the 2012 outburst using Swift/XRT data. Jet X-ray fluxes 
are found to be $1.195\times10^{-10}$, and $1.493\times10^{-10}$ $ergs/cm^2/s$ on MJD=56006 and 56013 respectively.
The jet X-ray fluxes were lower compared to the 2011 outburst.
 
MAXI~J1836-194 showed a strong jet activity during its 2011 outburst as compared to Swift~J1753.5-0127 during its 2005 
outburst. When the jet was the strongest in the HIMS of the 2011 outburst, its contribution in X-ray is found to be up to 
$\sim 86\%$. On an average, the jet X-ray contributed $\sim 41\%$ throughout the 2011 outburst. In the 2012 outburst, its 
contribution is found to be $\sim 54\%$ and $\sim 63\%$ in the two observations we studied, although total X-ray fluxes 
are much lower than that of the 2011 outburst. 

\subsubsection{\textbf{Radio and X-ray Flux Correlations}}

Standard jet models predict a correlation between the radio and the X-ray luminosity (Falcke \& Biermann 1995; 
Heinz \& Sunyaev 2003; Markoff et al. 2003; Russell et al. 2013). It was first observed for BHC GX~339-4 (Hannikainen
et al. 1998). Several BHCs show a standard correlation $F_R \sim F_X^b$ in the hard state, with a correlation index
of $b \sim 0.5-0.7$ (Corbel et al. 2003, 2013; Gallo et al. 2003; 2004; 2006). The correlation has been extended to 
the quiescent state (Gallo et al. 2014, Plotkin et al. 2013; 2017). The correlation is extended to AGNs in so-called 
`fundamental plane of black hole activity' by including the mass and the accretion rate (Merloni et al. 2003; Heinz 2004; 
Kording et al. 2006). However, some BHCs show more steeper index $\sim1.4$ (Coriat et al. 2011, Jonker et al. 2012). 
Other BHCs show dual correlation tracks. H~1743-322, XTE~J1752-223, MAXI~J1659-152 are some sources, where steeper
correlation indices are observed when $L_X$ \textgreater $10^{36}$ $ergs/s$. However, they are found to move towards 
the standard correlation track in the low luminosity state (Jonker et al. 2010, 2012; Coriat et al. 2011; Ratti et al.
2012). However, Swift~J1753.5-0127 has shown a different correlation index ($b \sim 1$) (Soleri et al. 2010, Rushtan 
et al. 2016, JCD17).

Interestingly, these correlations of $F_R$ have been obtained with the total X-ray flux ($F_X$) in $3-9$~keV band, 
which contains X-rays fluxes both from the disk and the jet. It may be the reason for different correlation indices 
in different spectral states of the same source. Since we have been able to separate the X-ray contributions of the
jet ($F_{ouf}$) and the accretion disk ($F_{inf}$) from total X-rays, the correlation between $F_R$ with $F_{ouf}$,
and $F_{inf}$ could be plotted. In Fig. 4a, a correlation plot between the $F_R$ vs. $F_{ouf}$ (in $3-25$~keV) is 
plotted, where the correlation index is found to be $b=0.61\pm0.08$, which is within the limit of standard correlation.
Here, we have used $7.45$~GHz VLA flux as $F_R$. For Swift~J1753.5-0127, a similar correlation index ($b=0.59\pm0.11$)
between $F_R$ vs. $F_{ouf}$ was also found (see, JCD17 for more details). This indicates that the mechanism for jet 
production could be same at least for these two BHCs. 

We have also drawn a plot of $F_R$ $vs$ $F_X$ for MAXI~J1836-194 during its 2011 outburst in Fig. 4(c-d) and it is 
fitted with $F_R \sim F_X^{b}$ to find the correlation index. For $3-25$ $keV$ PCU2 flux ($F_X$), we find
$b\sim 1.79\pm0.11$. Similarly, when we use $3-9$~keV PCU2 flux ($F_X$), $b\sim1.82\pm0.12$. Russell et al. (2015)
have also found a similar steeper correlation index ($b\sim1.8\pm0.2$). Interestingly, we do not find any correlation
between $F_R$ and $F_{inf}$ (see, Fig. 4b).

In Fig. 5, we draw the correlation plot in the luminosity i.e., $L_R - L_X$ plane. Other than the 2011 outburst of
MAXI~J1836-194, data from the 2012 outburst and the quiescent state between two outbursts are also included. 
We only show the variation of the total X-rays ($L_X$ in $2-10$~keV) with $L_R$ for the 2012 outburst and quiescent state.
For the 2011 outburst of MAXI~J1836-194 and the 2005 outburst of Swift~J1753.5-0127, we use TCAF
model fitted spectral result of RXTE/PCA from Paper I and JCD17 respectively. For the quiescent state as well as for the
2012 outburst of MAXI~J1836-194, we use $2-10$~keV MAXI/GSC data for calculating $L_X$ values, where radio data are 
available. It is to be noted that two data points from the quiescent state and one data point from the 2012 outburst are
not actual detection but they are the upper limits of the radio flux (Russell et al. 2014b, 2015). We also calculate the
jet X-ray luminosity ($L_{ouf}$) for MAXI~J1836-194 (during its 2011 outburst) as well as for Swift~J1753.5-0127 (during
its 2005 outburst) using TCAF model fitted PCA spectra. We find that the quiescent and the 2012 outburst data ($L_X$) 
points lie on the correlation track of jet ($L_R-L_{ouf}$) instead of their total X-ray track ($L_R-L_X$). This indicates 
that there was very little contribution in the X-rays from the accretion disk or inflowing matter. Only jet contributes
in the X-rays during the 2012 outburst as well as in the quiescent state. However, three upper limit points indicate that
actual radio fluxes could be at lower level and may fall close to the $L_R-L_X$ line.

In the case of compact jets, we expect a tight correlation between $F_R$ and $F_{ouf}$. For discrete ejections, a tight 
correlation may not be found. For BHC Swift~J1753.5-0127, a tight correlation is not obtained specially in the HIMS when
flux was much higher. In the HIMS, the nature of the jet is not entirely compact and perhaps partially blobby (see, JCD17 
and references therein). For MAXI~J1836-194, a tight correlation is obtained for $F_R$ and $F_{ouf}$ during both HS, and 
HIMS. This indicates that the nature of the jet is compact for this BHC, though the possibility of fast and slow components 
cannot be ruled out. Russell et al. (2015) also reported the compact nature of the jet for this BHC. 

\section{Discussions and Concluding Remarks}

Using the fact that the TCAF model normalization is a constant for a given source and observing instrument,
fits of good quality data is expected to have constant normalization. Any significant deviation indicates the presence
of other physical processes such as jet, disk precession etc. which are not included in TCAF based fits file used here.
Since the base of the jet/outflow also contributes to the X-ray along with the accretion disk, in a jet dominated phase,
a higher value of normalization is expected. This allows us to segregate the contributions from the inflow and the 
outflow components. During the 2011 outburst of MAXI~J1836-194, $N$ is generally varied in between $0.25-0.35$, except 
for the days with jet domination, when much higher $N$ values are required to fit the spectra satisfactorily. As in JCD17,
here we also assume that only accretion disk or inflowing matter contributes in the X-rays when the minimum value of 
$N= 0.25$ was required (as on 2011 Oct. 22, MJD=55856) while fitting $3-25$ $keV$ RXTE/PCA data of MAXI~J1836 during its
2011 outburst. Though the radio jet was appeared to be weakly active on this day, we assumed that jets contribution 
to the X-ray was negligible. To estimate the X-ray flux contributions coming only for the accretion disk or 
inflowing matter ($F_{inf}$), we refit all the spectra with frozen value of $N$ at $0.25$. Using Eq. 1, we calculate 
the jet contributions in X-rays ($F_{ouf}$). 

Chakrabarti (1999a,b) and Das \& Chakrabarti (1999) predicted that the ratio of the outflow rate to inflow rate is 
maximum in the HIMS when the shock strength is intermediate. In the present case also we observed maximum jet X-ray 
flux ($F_{ouf}$) in the HIMS for MAXI~J1836-194, which is similar to 2005 outburst of Swift~J1753.5-0127 (JCD17). 
MAXI~J1836-194 showed very strong jets in the X-rays. On 2011 September 15 (MJD=55819), almost $\sim 86\%$ X-ray flux 
came from the base of the jet. We find the X-ray jet luminosity $\sim5\times10^{36}$~$ergs/s$ on this date. After 
that, as the outburst progressed, both the total flux ($F_X$) and the jet flux ($F_{ouf}$) are decreased. 

Jet kinetic power is converted to radiations in different wavebands (radio, IR, optical to X-ray). Russell et al. 
(2014b) showed that there is a significant contribution in the jet from IR and optical wavebands which is presumably
due to synchrotron emission. However, since there is no real boundary between the CENBOL and the base of the jet,
in $F_{ouf}$, dominating contribution is due to inverse Comptonization and not due to synchrotron process. 

Russell et al. (2014b) estimated jet luminosity ($L_{jet}$) by integrating the jet spectra over $5 \times 10^9 - 7 
\times 10^{14}$ Hz for six observations during the 2011 outburst. They found $L_{jet}$ was minimum in the HIMS though
X-ray flux ($F_{ouf}$) was higher. This indicates that the jet was active in the X-ray but not in the IR and optical.
In the decay phase, the X-ray jet ($F_{ouf}$) was found to decrease. This is expected since the accretion rate also 
decreased which in turns reduced the mass outflow rate and jet X-ray flux ($F_{ouf}$). This agrees with the theoretical
point of view (C99a,b; Das \& Chakrabarti 1999). However, Russell et al. (2014b) found $L_{jet}$ increased in the decay
phase. Radio flux was lower in this phase. Thus most of the jet power was emitted in the IR and optical wavebands with
very little contribution in radio and X-ray (see Fig. 2 of this paper and Fig. 2 of Russell et al. 2014b). In the decay
phase, magnetic field at the base of the jet increased (Russell et al. 2014b). Due to high magnetic field, significant
amount of jet power was emitted in the IR and optical wavebands via synchrotron process. Since the jet is compact and 
dense, radio flux was very low due to high magnetic field. Thus, $L_{jet}$ increased due to high IR and optical radiation 
while radio and jet X-ray fluxes were lower.

After $\sim 3$ months of activity during the 2011 outburst, MAXI~J1836-194 entered in the quiescent state, which continued 
for the next $\sim4$ months. We calculated hardness ratio (HR) from $2-10$ $keV$ MAXI/GSC and $15-50$ keV Swift/BAT lightcurves.
HR is found to vary in between $2-6$, indicating the source was in the harder states. $3$ ATCA observations were available 
at $5.5$~GHz during the quiescent phase. However, radio flux densities were very low ($\sim 0.1$ $mJy$). On the 
contrary, during the 2011 outburst, in the HIMS, $5.5$~GHz ATCA flux density was $\sim 40$ $mJy$. X-ray luminosity ($L_X$) 
was also very low ($\sim10^{33}$) during this phase. Radio jet is generally observed in the hard and intermediate spectral
states. Observation of the radio jets and HR indicate that the source was in the low luminous HS during this phase of
the outburst. However, there are some reports that the quiescent state spectra is softer than that of the hard state
for several other BHCs, such as XTE~J1118+480 ($\Gamma\sim2.02$), XTE~J1550-564 ($\Gamma\sim2.25$), GX~339-4 ($\Gamma\sim 1.99$), 
V~404~Cyg ($\Gamma \sim 2.08$) (Corbel et al. 2006; Plotkin et al. 2013; 2017).

MAXI~J1836-194 also showed new flaring activity on 2012 March 12 (MJD=55998), which continued for $\sim 2$ months. We analyzed
$0.5-10.0$ $keV$ Swift/XRT spectra with TCAF model and extracted accretion flow parameters for the two observations from 
MJD=56006.85 \& 56013.05. Our model parameters indicate the high dominance of the sub-Keplerian halo accretion rates and high 
shock strengths during these observations (see, Table 1). This allowed us to infer that during our observed days of the 2012
outburst, the source was also in the HS. We also calculated jet contribution in the X-rays and found that the contributions 
were up to $\sim54\%$ and $\sim63\%$ of total X-ray in these two observations.

We expect $F_{ouf}$ and $F_R$ to be well correlated since they are both originated from the jet. At the base of the jet, 
X-ray is emitted since it is the hottest region right above the CENBOL. As the jet moves away, it is cooled due to expansion. 
Hence, it emits in the other low energy wavebands, such as UV, IR and radio. The correlation between the disk and the jet 
X-rays should directly lead to correlations with other radiations from the jet. We have drawn a correlation plot between 
the jet X-ray flux in $3-25$~keV of RXTE/PCA and radio in $7.45$~GHz VLA in Fig. 4(a). We obtained the correlation in the 
form of $F_R \sim F_{ouf}^{0.61\pm0.08}$. For Swift~J1753.5-0127, a similar correlation of $F_R \sim F_{ouf}^{0.59\pm 0.11}$ 
is found. Thus $F_R \sim F_{ouf}^{0.6}$ could be a universal correlation, but to firmly establish it, we need more observations. 
Interestingly, Corbel et al. (2003, 2013), Gallo et al. (2003) have found the standard correlation index to be $\sim 0.6-0.7$
if the correlation is done between $F_R$ and $F_X$. However, a steeper index ($\sim 1-1.4$) has been found for many `radio-quiet'
BHCs (Jonker et al. 2012, Ratti et al. 2012). An unusual steep index is found for MAXI~J1836-194 during its 2011 outburst when
$F_R$ was correlated with $F_X$. When we use $7.45$~GHz VLA data as $F_R$ with $3-25$~keV RXTE/PCA flux as $F_X$, a steeper 
correlation is observed with index $b\sim1.79\pm0.11$. Similarly, when we use $3-9$~keV RXTE/PCA flux as $F_X$, the index is 
found at $\sim 1.82\pm0.12$. A similar correlation was also reported by Russell et al. (2015). They suggested that it could be
due to variable Lorentz factor throughout the outburst. Another possible reason behind these steeper correlation indices may be
due to the high supply rate of low viscous sub-Keplerian matter in the form of wind or accretion, since its a low orbital period 
($\le 4.9$ hours) binary system. So, the system must be very compact and the companion winds may surround the accretion disk.
Scatterings of photons emitted from the jets with this cloud of wind matter could steepen the correlation indices.

In the quiescent state, and the subsequent 2012 outburst, the $L_X - L_R$ correlation points lie in the $L_{ouf} - L_R$
correlation track, i.e., jet-line of MAXI~J1836-194 at lower $L_X$ values. If the jet production mechanism remains the 
same across different outbursts, we may conclude that there was very little X-ray emission from the accretion disk in the 
quiescence and the 2012 outburst. All or most of the X-rays come from the jet or outflow. If this is true, then 
$L_{X} \sim L_{ouf}$. Thus it is possible that some alternate mechanism may exist for producing jets and outflows in the
quiescent and low luminosity phases of these types of black hole binary systems.

During the 2011 outburst, the jet contribution is as high as $\sim 86\%$ of the total X-ray with an average contribution of 
$\sim 41\%$. This makes MAXI~J1836-194 as a jet dominated black hole candidate. Even after the 2012 outburst, ATCA observed 
radio flux densities of $0.07$~mJy at $5.5$~GHz and $0.09$~mJy at $9$~GHz on MJD=56163. This also indicates the activity of
jet on that day. Thus even when the X-ray luminosity is less than $\sim10^{32}$ $ergs/s$, the jet is active. Chakrabarti
\& Bhaskaran (1992) showed that it is easier to produce jet from the sub-Keplerian halo. In the quiescent state, Keplerian 
disk may not be formed due to the lack of viscosity. However, a constant supply of wind from the companion star form 
radiatively inefficient sub-Keplerian flow which launch the jets. This implies that in the quiescent state as well as during 
the 2012 outburst, the X-ray contribution from the accretion disk is very negligible.

$L_R - L_{ouf}$ correlation is expected to hold tight as long as the jet remains compact. In case of blobby jet or discrete 
ejection, we do not expect a tight correlation. For BHC Swift~J1753.5-0127, we observe scatter points in the $L_R - L_{ouf}$
track (see, Fig. 5) specially in the HIMS. Thus nature of the jet for Swift~J1753.5-0127 is blobby in the HIMS. However in HS,
when the fluxes are low, a tighter correlation is obtained. It indicates that the jet remains compact as in the HS of 
Swift~J1753.5-0127 during its 2005 outburst (JCD17). From Fig. 5, for MAXI~J1836-194, we obtain a tight correlation in both HS
and HIMS. It leads us to conclude that the jet could be compact throughout the outburst, though the possibility of a fast and 
slow components cannot be ruled out. Russell et al. (2015) also reported compact nature of the jet for this BHC. 

In future, we would like to find disk-jet coupling for other black hole candidates. So far, we found $F_R \sim F_{ouf}^{0.6}$
for MAXI~J1836-194 and Swift~J1753.5-0127. We would like to see if this relation holds in some other cases as well.

\section*{Acknowledgments}
A.J. and D.D. acknowledge support from DST/GITA sponsored India-Taiwan collaborative project fund (GITA/DST/TWN/P-76/2017).
D.C. and D.D. acknowledge support from DST/SERB sponsored Extra Mural Research project (EMR/2016/003918).
This research has made use of the XRT Data Analysis Software (XRTDAS) developed under the responsibility of the 
ASI Science Data Center (ASDC), Italy

{}


\clearpage

\begin{figure}
\vskip 0.0cm
\centerline{
\includegraphics[scale=0.6,angle=0,width=8.5truecm]{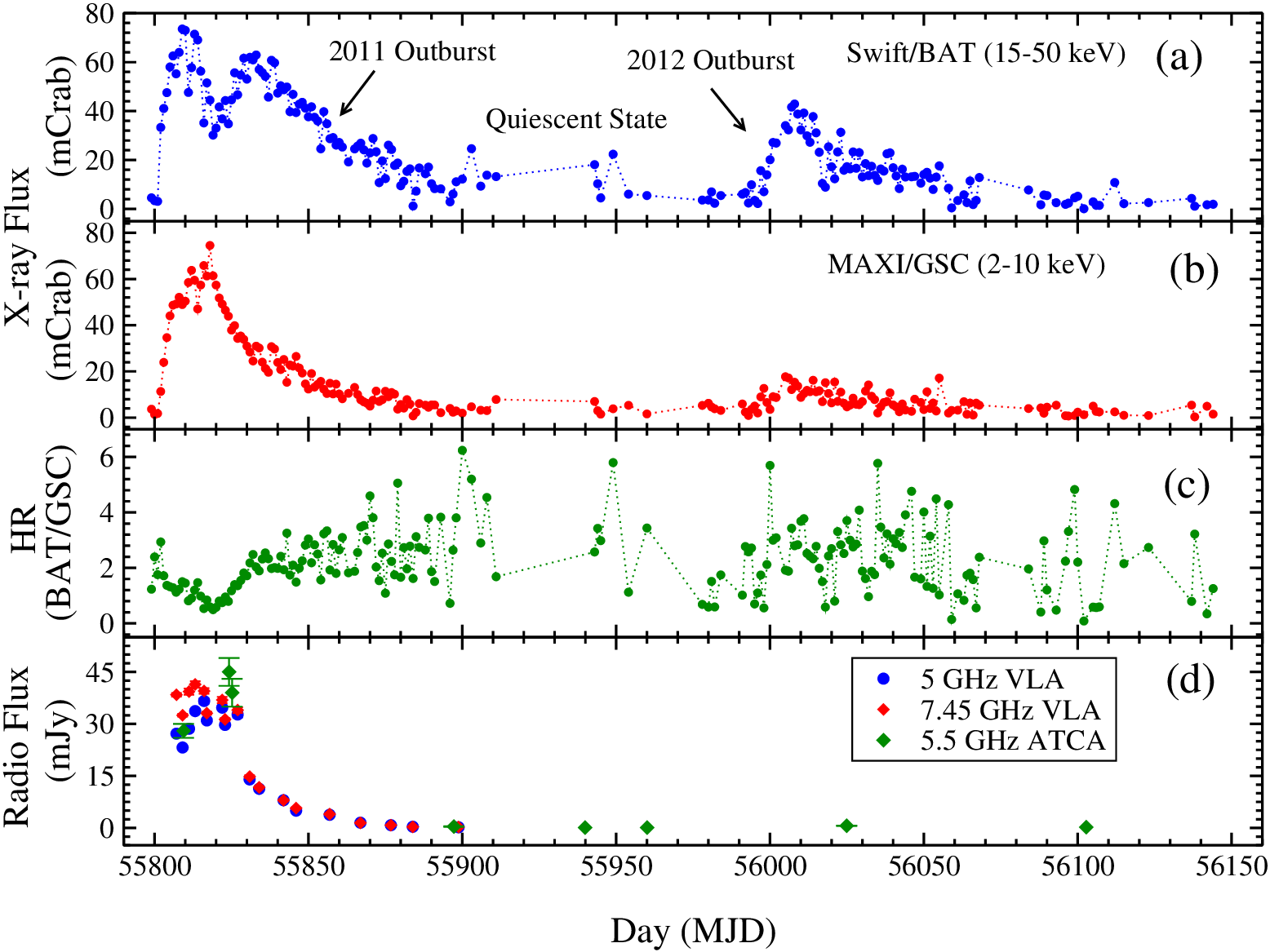}}
\label{fig1}
\caption{Variation of (a) $15-50$ $keV$ $Swift/BAT$ flux in mCrab, (b) $2-10$ $keV$ $MAXI/GSC$ flux in mCrab,
(c) Hardness ratio (HR = ($15-50$ $keV$)/($2-10$ $keV$) rates), and (d) $5$~GHz VLA, $5.5$~GHz ATCA and $5.5$~GHz 
VLA radio flux densities in mJy unit are shown. Radio data are taken from Russell et al. (2015). The 2011
outburst begins on MJD = 55804 and continued for $\sim 3$~months till MJD $\sim 55890$. The 2012 outburst begin on 
MJD=56000 and remain active for $\sim2$ months. In between 2011 and 2012 outbursts, the source was in quiescent state 
for $\sim 4$ months.}
\end{figure}

\begin{figure}
\vskip 0.0cm
\centerline{
\includegraphics[scale=0.6,angle=0,width=8.5truecm]{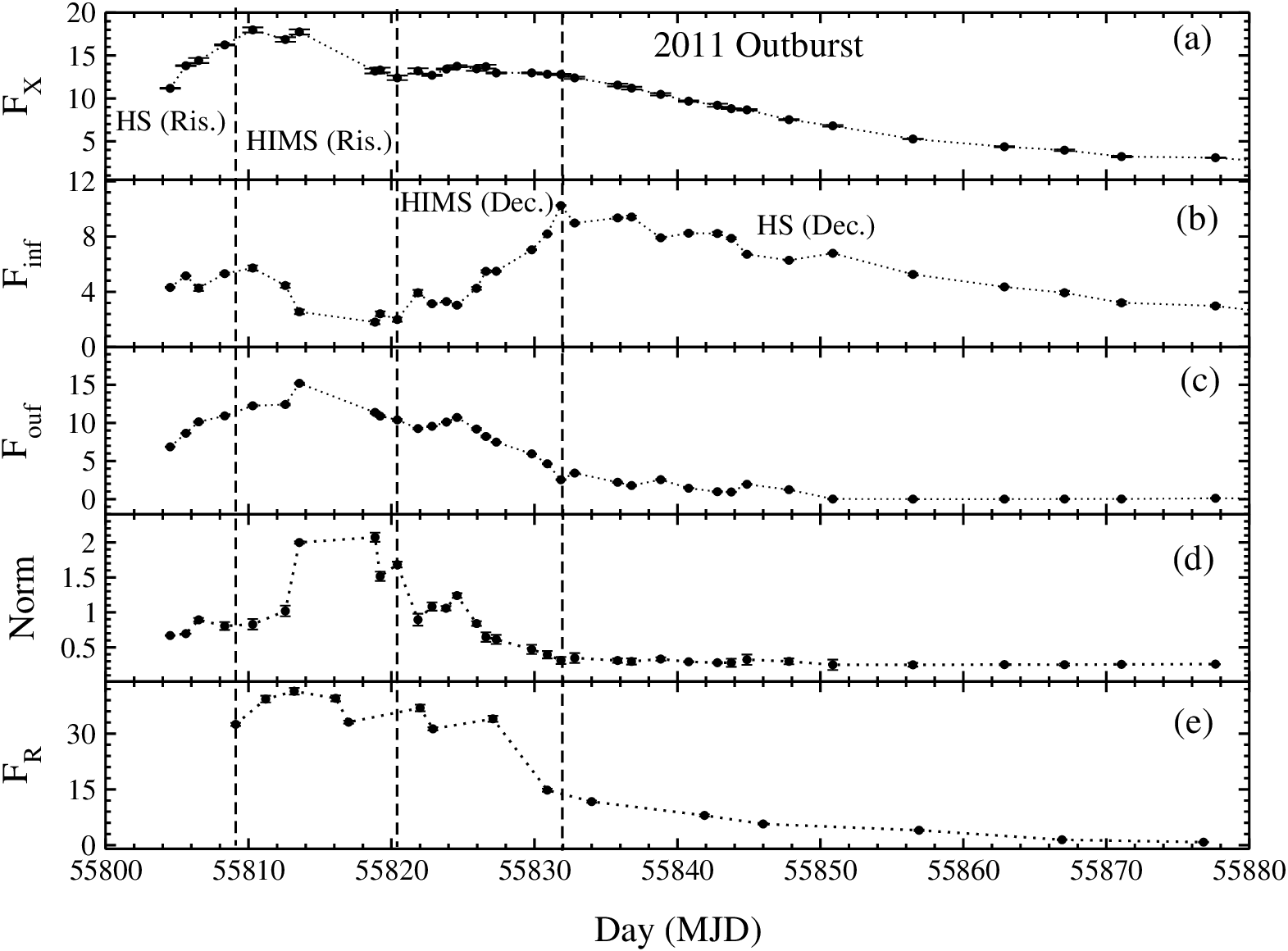}}
\label{fig2}
\caption{Variation of (a) total X-ray flux ($F_X$), (b) the X-ray flux from accretion disk ($F_{inf}$),
(c) the jet X-ray flux ($F_{ouf}$), (d) TCAF normalization ($N$), and (e) $7.45$~GHz Radio flux density ($F_R$) of
VLA observation are shown with day (MJD) during the 2011 outburst. The radio data are taken from Russell et al. 
(2014b, 2015). X-ray flux are in the RXTE/PCA energy range of $3-25$~keV and in the unit of $10^{-9}$ $ergs/cm^2/s$.}
\end{figure}

\begin{figure}
\vskip 0.0cm
\centerline{
\includegraphics[scale=0.6,angle=0,width=8.5truecm]{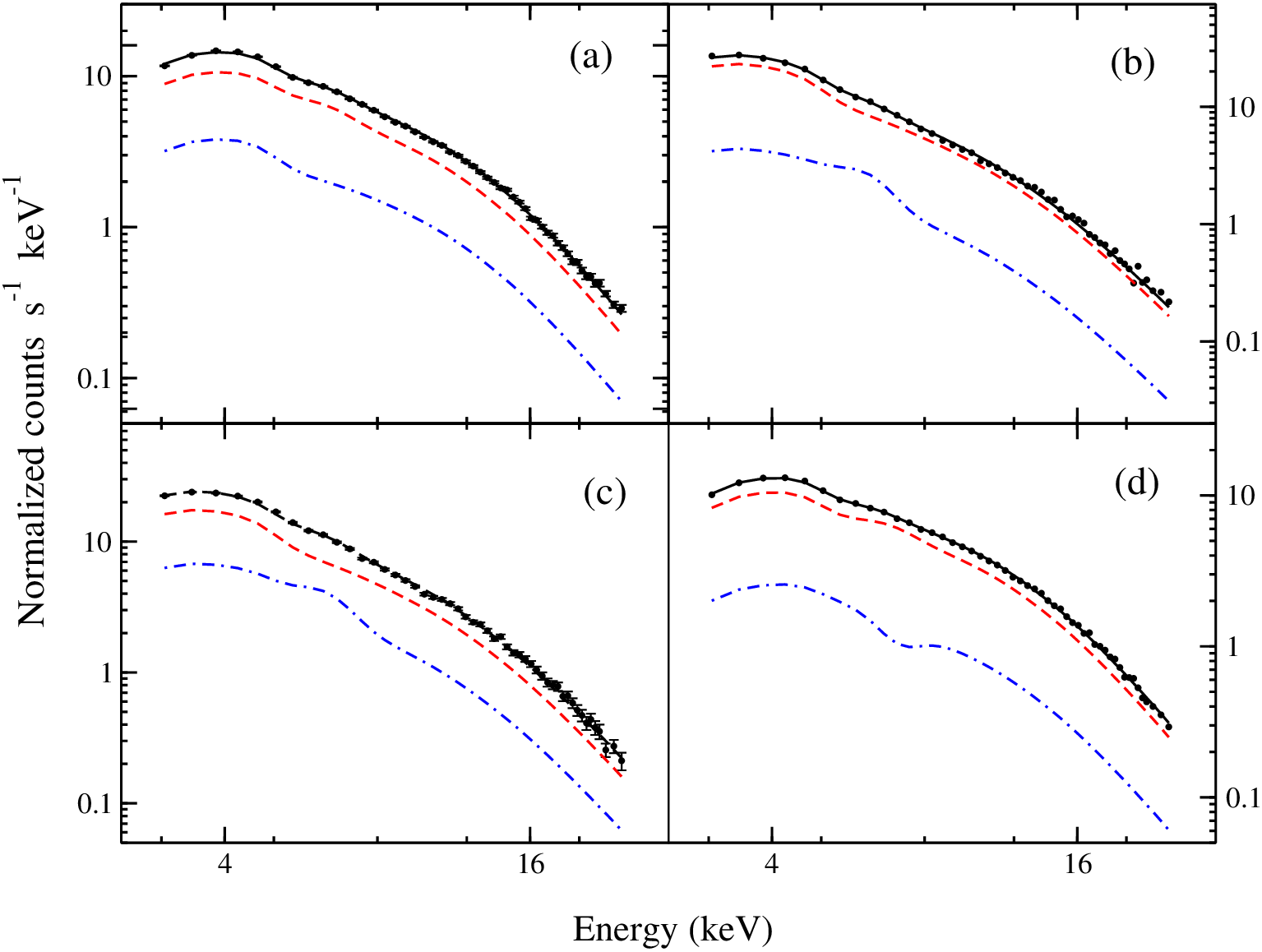}}
\label{fig3}
\caption{Four spectra selected from four different spectral states, fitted with the TCAF model fits file 
by keeping model normalization as free (blue curve) or frozen at N=0.25 (red curve). The spectra are from 
observation IDs: (a) 96371-03-01-00 (HS - Ris.), (b) 96438-01-01-04 (HIMS - Ris.), (c) 96438-01-02-00 (HIMS - Dec.),
and (d) 96438-01-04-01 (HS - Dec.). Jet X-ray spectra are shown in blue dot-dashed curves. }
\end{figure}

\begin{figure}
\vskip 0.0cm
\centerline{
\includegraphics[scale=0.6,angle=0,width=8.5truecm]{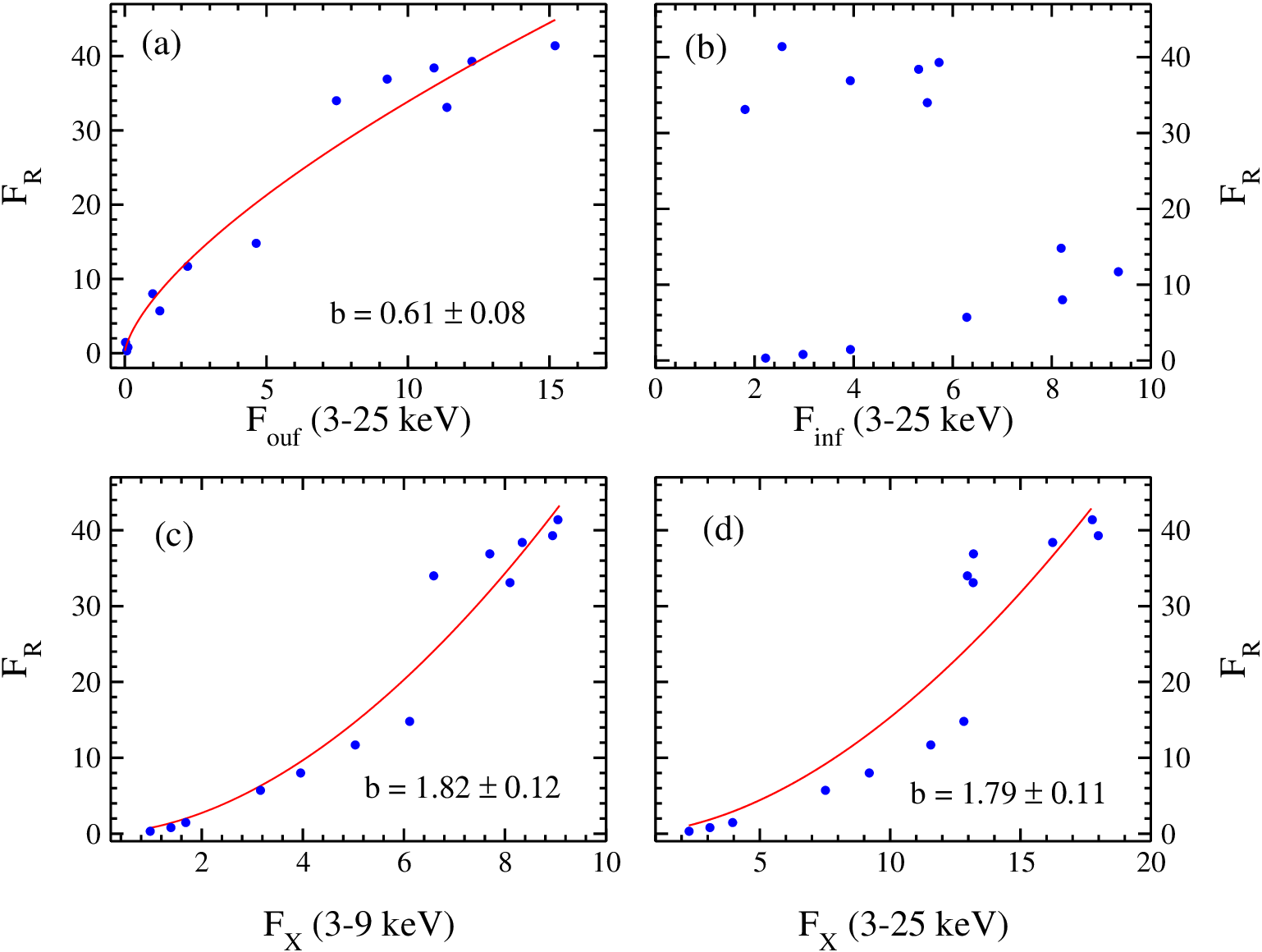}}
\label{fig4}
\caption{Correlation plots are shown for $7.45$~GHz $F_R$ with (a) $3-25$~keV $F_{ouf}$, (b) $3-25$~keV $F_{inf}$, 
(c) $3-9$~keV $F_{X}$ and (d) $3-25$ $keV$ $F_{X}$. The correlations are in the form of $F_R \sim F_X^b$, where 
$b$ is the correlation index. We did not find any correlation between $F_R$ and $F_{inf}$. $F_R$ is in the 
unit of $mJy$. $F_X$, $F_{ouf}$ and $F_{inf}$ are in the unit of $10^{-10}$ $ergs/cm^2/s$.}
\end{figure}

\begin{figure}
\vskip 0.0cm
\centerline{
\includegraphics[scale=0.6,angle=0,width=8.5truecm]{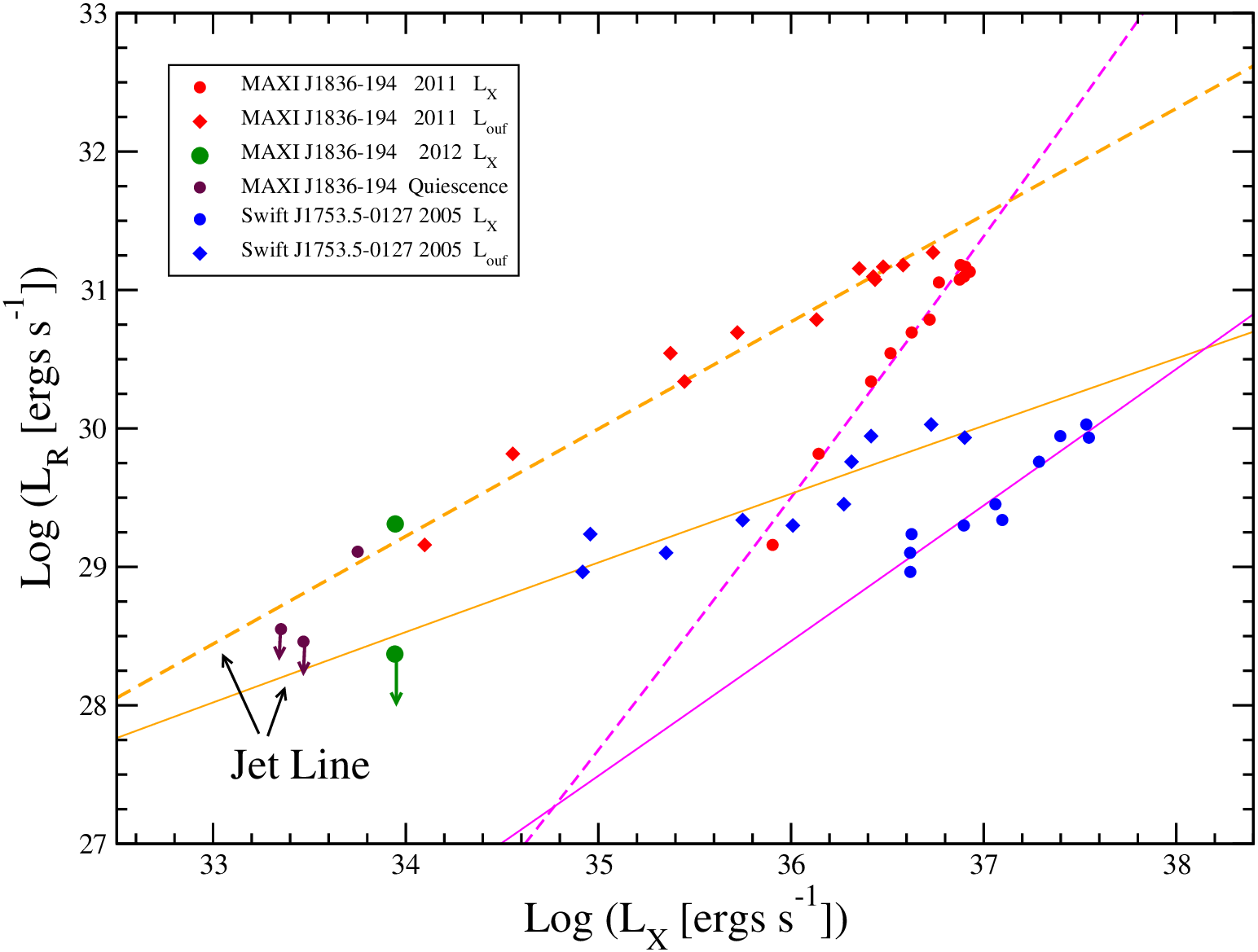}}
\label{fig5}
\caption{Correlation plots are shown in luminosity plane. Orange dashed and solid lines represent correlation track 
of $L_R - L_{ouf}$ for MAXI~J1836-194 during the 2011 outburst and Swift~J1753.5-0127 during the 2005 outburst respectively. 
Red and blue diamonds represent $L_R - L_{ouf}$ points for MAXI~J1836-194 and Swift~J1753.5-0127 respectively. 
Red and blue circles represent $L_R - L_X$ correlation points for MAXI~J1836-194 during the 2011 outburst and 
Swift~J1753.5-0127 during the 2005 outburst respectively. Purple dashed and solid lines are for $L_R - L_X$ correlation 
track for MAXI~J1836-194 during the 2011 outburst and Swift~J1753.5-0127 during the 2005 outburst respectively. 
Green and brown circle represent $L_R - L_X$ points for MAXI~J1836-194 during the 2012 outburst and quiescent state 
respectively. Note: X-ray luminosities ($L_X$) are calculated in $2-10$~keV range of PCA \& XRT by assuming distance
of MAXI~J1836-194 and Swift~J1753.5-0127 as $7$~kpc and $8$~kpc respectively. 'Down arrow' indicate points with 
upper limit of the radio luminosities.}
\end{figure}

\clearpage

\begin{center}
\vskip -2.0cm
\addtolength{\tabcolsep}{-4.50pt}
\scriptsize
\centering
\centering{\large \bf Table I}
\vskip 0.2cm
\centerline {Swift/XRT results fitted with the TCAF solution}
\vskip 0.2cm

\begin{tabular}{|c|c|c|c|c|c|c|c|c| } 
 \hline
 Obs ID & MJD & $\dot{m_d}$ & $\dot{m_h}$ & $X_s$ & $R$ & $N$ & Line E & $\chi^2 /dof$  \\
  & (Day)  & ($\dot{M}_{Edd}$) & ($\dot{M}_{Edd}$) & ($r_s$) & & & ($keV$) & \\
 (1) & (2) & (3) & (4) & (5) & (6) & (7) & (8) & (9)  \\
 \hline
 00032087028 & 55859.21 &  $1.698^{\pm 0.023}$ &$0.584^{\pm 0.013}$ &$73.19^{\pm 2.44}$  &$1.081^{\pm 0.087}$ &$34.21^{\pm 1.93}$ &$6.33^{\pm 0.22}$ & 833.2/940 \\
\hline
 00032308002 & 56006.85 &  $0.011^{\pm 0.001}$ &$0.162^{\pm 0.002}$ &$281.22^{\pm 5.65}$ &$3.567^{\pm 0.145}$ &$74.92^{\pm 2.23}$ &$6.68^{\pm 0.19}$ & 760.3/940 \\
 00032308005 & 56013.05 &  $0.012^{\pm 0.001}$ &$0.158^{\pm 0.002}$ &$271.24^{\pm 6.22}$ &$3.557^{\pm 0.129}$ &$93.82^{\pm 1.91}$ &$6.61^{\pm 0.14}$ & 755.7/940 \\ 
 \hline
\end{tabular}
\noindent{
\leftline{First observation is from the 2011 outburst. Last two observations are from the 2012 outburst. Data analysis are done}
\leftline{using $0.5-10.0$~keV Swift/XRT data with TCAF solution. Accretion rates ($\dot{m_d}$ \& $\dot{m_h}$) are represented
in Eddington}
\leftline{accretion rate ($\dot{M}_{Edd}$). shock location ($X_s$) is presented in Schwarzschild radius ($r_s$). $R$ is the
compression ratio (ratio }
\leftline{of post-shock density to pre-shock density). Mass is kept frozen at $9.54$ $M_{\odot}$.}
\leftline{$N$ is TCAF normalization. Gaussian `line E' represent peak energy of $Fe-k_{\alpha}$ line.}
}
\end{center}


\begin{center}
\vskip 1.0cm
\addtolength{\tabcolsep}{-4.50pt}
\scriptsize
\centering
\centering{\large \bf Table II}
\vskip 0.2cm
\centerline {X-ray jet during 2011 and 2012 outburst}
\vskip 0.2cm
\begin{tabular}{|c|c|c|c|c|c|c|c|}
\hline
 No. & Obs ID & MJD & $N$ & $F_X$ & $F_{inf}$ & $F_{ouf}$ & \% of $F_{ouf}$ \\
 (1) & (2) & (3) & (4) & (5) & (6) & (7) & (8) \\
 \hline
1 &X-01-00&$ 55804.52$&$ 0.669^{\pm 0.002}$& $  11.181^{\pm 0.052}$&$  4.318^{\pm 0.041}$&$  6.863^{\pm 0.011}$&$ 61.38$\\
2 &X-02-00&$ 55805.61$&$ 0.695^{\pm 0.002}$& $  13.811^{\pm 0.049}$&$  5.159^{\pm 0.035}$&$  8.652^{\pm 0.014}$&$ 62.64$\\
3 &X-03-00&$ 55806.51$&$ 0.893^{\pm 0.027}$& $  14.417^{\pm 0.290}$&$  4.276^{\pm 0.212}$&$ 10.141^{\pm 0.078}$&$ 70.34$\\
4 &X-03-01&$ 55808.33$&$ 0.804^{\pm 0.057}$& $  16.235^{\pm 0.057}$&$  5.313^{\pm 0.071}$&$ 10.922^{\pm 0.014}$&$ 67.27$\\
5 &X-03-02&$ 55810.29$&$ 0.829^{\pm 0.076}$& $  17.985^{\pm 0.301}$&$  5.725^{\pm 0.192}$&$ 12.260^{\pm 0.104}$&$ 68.16$\\
6 &X-03-03&$ 55812.57$&$ 1.020^{\pm 0.077}$& $  16.865^{\pm 0.271}$&$  4.451^{\pm 0.156}$&$ 12.414^{\pm 0.096}$&$ 73.60$\\
7 &Y-01-00&$ 55813.55$&$ 1.998^{\pm 0.017}$& $  17.755^{\pm 0.299}$&$  2.554^{\pm 0.162}$&$ 15.201^{\pm 0.077}$&$ 85.61$\\
8 &Y-01-04&$ 55818.84$&$ 2.073^{\pm 0.066}$& $  13.185^{\pm 0.281}$&$  1.807^{\pm 0.147}$&$ 11.378^{\pm 0.134}$&$ 86.29$\\
9 &Y-01-05&$ 55819.20$&$ 1.515^{\pm 0.067}$& $  13.308^{\pm 0.283}$&$  2.413^{\pm 0.167}$&$ 10.895^{\pm 0.116}$&$ 81.86$\\
10&Y-02-03&$ 55820.40$&$ 1.683^{\pm 0.042}$& $  12.394^{\pm 0.271}$&$  1.993^{\pm 0.152}$&$ 10.401^{\pm 0.119}$&$ 83.92$\\
11&Y-02-00&$ 55821.85$&$ 0.894^{\pm 0.085}$& $  13.198^{\pm 0.315}$&$  3.933^{\pm 0.209}$&$  9.265^{\pm 0.106}$&$ 70.19$\\
12&Y-02-01&$ 55822.83$&$ 1.083^{\pm 0.059}$& $  12.699^{\pm 0.078}$&$  3.139^{\pm 0.051}$&$  9.560^{\pm 0.027}$&$ 75.28$\\
13&Y-02-04&$ 55823.84$&$ 1.059^{\pm 0.031}$& $  13.416^{\pm 0.075}$&$  3.296^{\pm 0.042}$&$ 10.119^{\pm 0.029}$&$ 75.43$\\
14&Y-02-05&$ 55824.58$&$ 1.240^{\pm 0.033}$& $  13.751^{\pm 0.102}$&$  3.033^{\pm 0.054}$&$ 10.718^{\pm 0.056}$&$ 77.94$\\
15&Y-02-02&$ 55825.94$&$ 0.841^{\pm 0.033}$& $  13.453^{\pm 0.242}$&$  4.256^{\pm 0.102}$&$  9.197^{\pm 0.130}$&$ 68.36$\\
16&Y-02-06&$ 55826.60$&$ 0.647^{\pm 0.068}$& $  13.701^{\pm 0.223}$&$  5.481^{\pm 0.088}$&$  8.220^{\pm 0.112}$&$ 59.99$\\
17&Y-03-04&$ 55827.33$&$ 0.613^{\pm 0.065}$& $  12.962^{\pm 0.071}$&$  5.488^{\pm 0.050}$&$  7.474^{\pm 0.021}$&$ 57.66$\\
18&Y-03-01&$ 55829.80$&$ 0.470^{\pm 0.066}$& $  12.984^{\pm 0.061}$&$  7.043^{\pm 0.028}$&$  5.941^{\pm 0.033}$&$ 45.75$\\
19&Y-03-05&$ 55830.89$&$ 0.395^{\pm 0.053}$& $  12.826^{\pm 0.057}$&$  8.189^{\pm 0.024}$&$  4.637^{\pm 0.037}$&$ 36.15$\\
20&Y-03-02&$ 55831.84$&$ 0.313^{\pm 0.050}$& $  12.785^{\pm 0.091}$&$ 10.241^{\pm 0.035}$&$  2.544^{\pm 0.056}$&$ 19.89$\\
21&Y-03-03&$ 55832.81$&$ 0.348^{\pm 0.070}$& $  12.396^{\pm 0.122}$&$  8.976^{\pm 0.064}$&$  3.420^{\pm 0.058}$&$ 27.58$\\
22&Y-04-01&$ 55835.81$&$ 0.311^{\pm 0.031}$& $  11.557^{\pm 0.152}$&$  9.345^{\pm 0.078}$&$  2.212^{\pm 0.084}$&$ 19.13$\\
23&Y-04-02&$ 55836.78$&$ 0.298^{\pm 0.044}$& $  11.201^{\pm 0.154}$&$  9.417^{\pm 0.098}$&$  1.784^{\pm 0.096}$&$ 15.92$\\
24&Y-04-04&$ 55838.83$&$ 0.333^{\pm 0.025}$& $  10.473^{\pm 0.140}$&$  7.914^{\pm 0.071}$&$  2.559^{\pm 0.069}$&$ 24.43$\\
25&Y-04-06&$ 55840.78$&$ 0.294^{\pm 0.007}$& $   9.678^{\pm 0.079}$&$  8.246^{\pm 0.058}$&$  1.432^{\pm 0.021}$&$ 14.79$\\
26&Y-05-00&$ 55842.79$&$ 0.280^{\pm 0.008}$& $   9.200^{\pm 0.179}$&$  8.217^{\pm 0.118}$&$  0.982^{\pm 0.067}$&$ 10.67$\\
27&Y-05-01&$ 55843.76$&$ 0.280^{\pm 0.057}$& $   8.805^{\pm 0.045}$&$  7.869^{\pm 0.071}$&$  0.936^{\pm 0.026}$&$ 10.62$\\
28&Y-05-04&$ 55844.86$&$ 0.323^{\pm 0.074}$& $   8.666^{\pm 0.084}$&$  6.714^{\pm 0.037}$&$  1.952^{\pm 0.057}$&$ 22.52$\\
29&Y-05-06&$ 55847.80$&$ 0.300^{\pm 0.042}$& $   7.518^{\pm 0.069}$&$  6.283^{\pm 0.025}$&$  1.235^{\pm 0.044}$&$ 16.42$\\
30&Y-06-02&$ 55850.87$&$ 0.251^{\pm 0.074}$& $   6.805^{\pm 0.097}$&$  6.794^{\pm 0.041}$&$  0.011^{\pm 0.056}$&$  0.16$\\
31&Y-07-00&$ 55856.47$&$ 0.250^{\pm 0.027}$& $   5.263^{\pm 0.014}$&$  5.263^{\pm 0.087}$&$  0.000^{\pm 0.073}$&$  0.00$\\
32&Y-08-00&$ 55862.87$&$ 0.253^{\pm 0.011}$& $   4.367^{\pm 0.075}$&$  4.358^{\pm 0.024}$&$  0.008^{\pm 0.099}$&$  0.19$\\
33&Y-08-05&$ 55867.08$&$ 0.252^{\pm 0.016}$& $   3.956^{\pm 0.087}$&$  3.935^{\pm 0.132}$&$  0.021^{\pm 0.055}$&$  0.54$\\
34&Y-09-02&$ 55871.06$&$ 0.256^{\pm 0.007}$& $   3.216^{\pm 0.088}$&$  3.205^{\pm 0.125}$&$  0.010^{\pm 0.037}$&$  0.31$\\
35&Y-10-01&$ 55877.63$&$ 0.259^{\pm 0.002}$& $   3.085^{\pm 0.022}$&$  2.978^{\pm 0.077}$&$  0.107^{\pm 0.055}$&$  3.47$\\
\hline
36&$00032308002^*$& 56006.85&$74.92^{\pm 2.23}$&$2.201^{\pm 0.023}$&$1.101^{\pm 0.012}$ &$1.193^{\pm 0.011}$&$54.23$ \\
37&$00032308005^*$& 56013.05&$93.87^{\pm 1.99}$&$2.353^{\pm 0.024}$&$0.859^{\pm 0.010}$ &$1.493^{\pm 0.014}$&$63.48$ \\
\hline
\end{tabular}

\leftline {Note: $F_X$, $F_{ouf}$ and $F_{inf}$ are the unit of $10^{-10}$ $ergs/cm^2/s$.}
\leftline {For the 2011 outburst, $F_X$, $F_{ouf}$ and $F_{inf}$ are calculated in the energy range of $3-25$~keV.}
\leftline {For the 2012 outburst, fluxes are calculated in $0.5-10.0$~keV energy range. }
\leftline {$F_{ouf}/F_X$ indicate \% of jet X-ray contributions out of total X-rays.}
\leftline {X=96371-03, Y=96438-01 are prefixes of observation ID's. $^*$ are from the 2012 outburst. }
\end{center}

\end{document}